\documentclass[10pt]{iopart} 
\usepackage{graphicx}
\usepackage{cite} 
\usepackage{xcolor} 
\usepackage{soul}   
\begin{document}

\title[Revealing laser-coherent electron features using phase-of-the-phase spectroscopy]{Revealing laser-coherent electron features using phase-of-the-phase spectroscopy}

\author{V.A. Tulsky$^1$, B. Krebs$^1$, J. Tiggesb{\"a}umker$^{1,2}$, D. Bauer$^1$}

\address{$^1$ Institute of Physics, University of Rostock, 18051 Rostock, Germany}
\address{$^2$ Department Life, Light and Matter, University of Rostock,
18051 Rostock, Germany}



\begin{abstract}
  Phase-of-the-phase (PoP) spectroscopy is extended to two-color laser
  fields having a circularly counter-rotating polarization. In
  particular, the higher harmonics of the (two-color) phase
  information are analyzed in order to extract the laser-coherent part
  of the photoelectron spectra taken under complex target
  conditions. We illustrate this with a proof-of-principle simulation
  by considering strong-field electron emission from argon atoms
  within helium nanodroplets under realistic experimental conditions,
  i.e., a limited number of photoemission events. Multiple elastic
  scattering on neutral helium atoms creates a laser-incoherent
  background, but the higher harmonics of the PoP-signal
  allow to resolve the coherent contribution to
  the photoemission.
\end{abstract}

%
%
%
\maketitle

\ioptwocol

\section{Introduction}
Photoelectron momentum distributions obtained in experiments with
strong laser fields are a treasury containing valuable information
related to the target, the laser field, and the ultrafast phenomena
during the process of ionization. However, in complex targets
scattering  may destroy
the phase relation between the photoelectron yield and the laser
field, and these electrons may clutter the
coherent electron spectral features known from atomic
gas-phase targets (such as plateaus, low-energy structures etc.). The
recently proposed phase-of-the-phase (PoP) technique
\cite{Skruszewicz_PhysRevLett_2015} can serve as Occam's razor with
respect to incoherent processes that spoil photoelectron spectra
(PES).  The general principle of the PoP and similar techniques is based on the irradiation of a target with a laser field that has a periodic
parameter such as the relative phase between two fields with different
colors \cite{Skruszewicz_PhysRevLett_2015, Almajid_JPhysB_2017, Wurzler_JPhysB_2017, Beaulieu_Science_2017, Tan_OptQuantEl_2018, Han_PhysRevLett_2018, Porat_NatComm_2018, Tan_OptExpr_2018, Tulsky_PhysRevA_2018_POP} or the carrier-envelope phase \cite{Rupp_JModOpt_2017}. However, in contrast to techniques where the forward-backward asymmetry \cite{Seiffert_JPhys_B_2017, Gao_PhysRevA_2017, Luo_PhysRevA_2017, Kubel_MolPhys_2017, Azoury_NatComm_2017, Zhang_OptLett_2017} or side streaking \cite{Eicke_JPhysB_2017, Gong_PhysRevLett_2017, Xie_PhysRevLett_2017, Yu_PhysRevA_2018, Richter_PhysRevLett_2015, Richter_PhysRevA_2016} of the PES is analyzed at selected values of the periodic parameter, the PoP suggests to perform the Fourier transform with respect to this periodic parameter and to reveal how the PES follow it, e.g., whether the PES change with the same periodicity as the relative phase (``first harmonic'') but a certain phase lag $\Phi_1$ (the ``phase of the phase''), twice per period (``second harmonic'', with the corresponding phase lag $\Phi_2$), and so on.  It turns out that these Fourier components have well distinguished features at certain momenta. Specifically, the phase lag of some momentum-resolved Fourier components sharply flips along certain curves in momentum-space. Those ``flipping curves'' are sensitive to the parameters of the system, i.e., the laser intensity and frequency, and the type of target used.

In Ref.~\cite{Tulsky_PhysRevA_2018_POP}, the PoP technique
was developed for the case of two-color circularly polarized
counter-rotating intense laser fields with frequency ratio 1:2. Here,
we extend the method to arbitrary ratios. We establish the general property of
the PoP flipping curves, showing that they all have a quite simple,
circular geometry and therefore are easier to analyze than, for
instance, the features observed in the PES for two-color linearly
polarized fields \cite{Skruszewicz_PhysRevLett_2015,
Almajid_JPhysB_2017}. One of the flipping curves was predicted in
\cite{Tulsky_PhysRevA_2018_POP}. 
However, for fast photoelectrons, where the signals are low as a consequence of the significantly reduced ionization probability, flipping curves may not be resolved due to the small number of events on the detector.
  In the present paper, we propose a way
to bypass this problem extending the theory so that it can be used for
a broader domain of parameters. We choose the specific frequency ratio
$\omega$-$2\omega$ in most of our examples, as it is widely used in both
experimental and theoretical investigations
\cite{Mancuso_PhysRevA_2015, Eckart_PhysRevLett_2016, Mancuso_PhysRevA_2016, Baykusheva_PhysRevLett_2016, Mancuso_PhysRevA_2017, Ayuso_JPhysB_2018, Jimenez_2018_PhysRevA}. Note,
however, that for other frequency ratios $a\omega$-$b\omega$, being of
particular interest nowadays (see, e.g., \cite{Mauger_JPhysB_2016, Milosevic_PhysRevA_2016, Habibovic_OptQuantEl_2018, Milosevic_atoms_2018, Katsoulis_arxiv_2019}), the PoP technique
can also be applied.

The paper is organized as follows. Section 2 is devoted to the
derivation of the relevant formulas within the Strong Field
Approximation (SFA) \cite{Keldysh_1965, Faisal_1973, Reiss_1980}, and
the general properties of PoP spectra are derived. In Section~3, the
PES obtained for a complex target where multiple elastic scattering of
electrons after ionization are taken into account is presented. Elastic
scattering gives rise to a huge contribution of the incoherent
electron signal (with respect to the phase shift between the laser
field components). The PoP technique is then applied in order to
reveal the small coherent features in the electron yield. We conclude in Section~4. Atomic units are used
throughout this paper unless otherwise stated.
\begin{center}
	\begin{figure*}
	\includegraphics[scale=0.098]{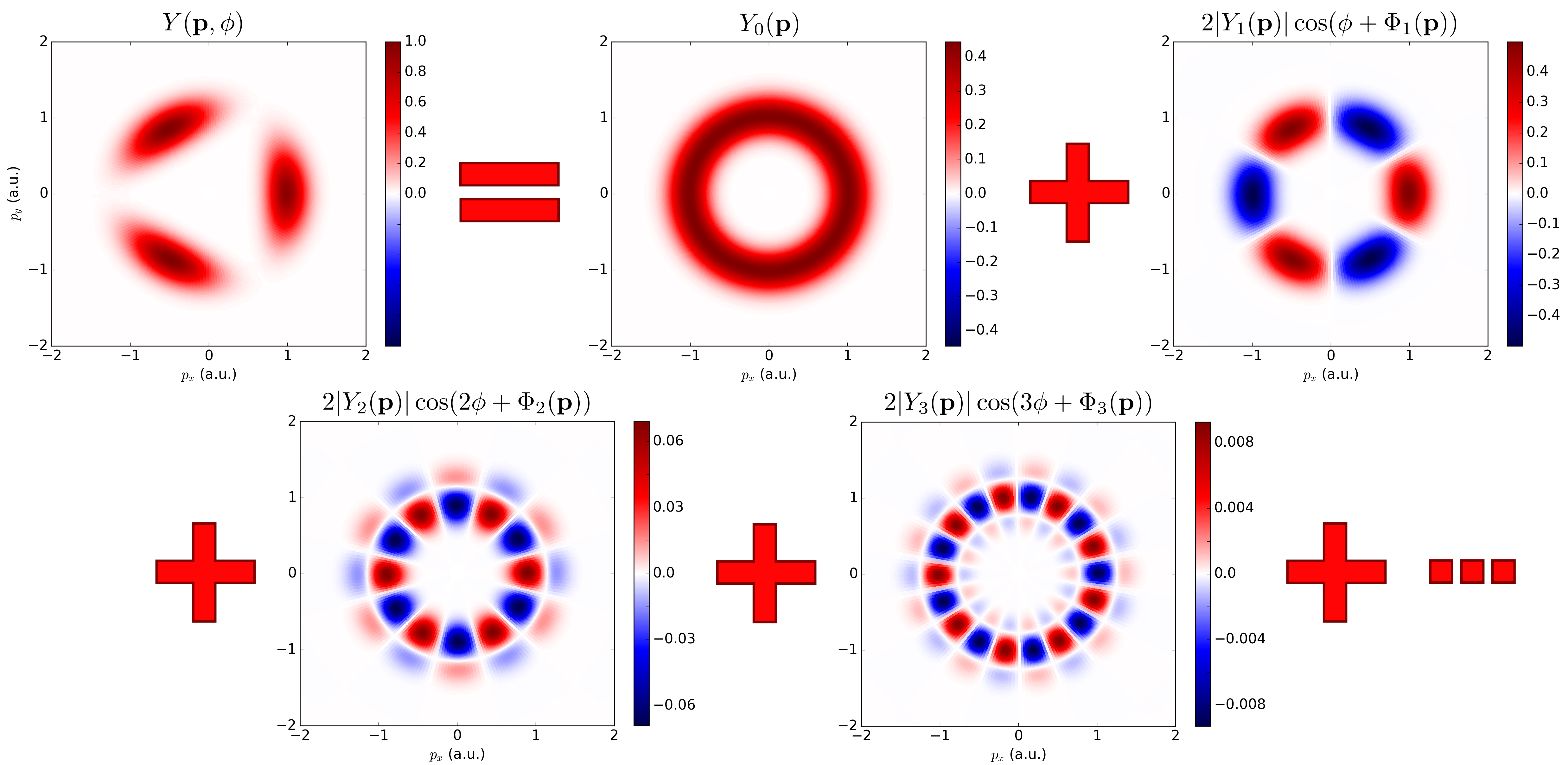}
	\caption{Decomposition of a total photoelectron spectrum into
Fourier components with respect to the phase shift between the two
colors of the laser field (calculated according to (\ref{SFA_Yield})
with interference terms discarded). The laser field defined by
(\ref{A_t}) with $a=1, b=2$, intensity and wavelength of the main
component of the laser $I = 2 \cdot10^{14}$ W/cm$^2$, $\lambda_1 =
800$\,nm, $\xi=0.05$, $\phi=0$, and ionization potential
$I_p=15.8$\,eV (argon).}
	\label{fig:fig1}
	\end{figure*}
\end{center}

\section{Theory}
\subsection{Strong field approximation} Let the vector potential of a
laser field in dipole approximation be of the form 
\begin{equation}
{\bf A}(t) = {\bf A}_a(a\omega t) + {\bf A}_b(b\omega t +
\phi).
\end{equation} 
The PoP technique is based on the
momentum-resolved PES, Fourier-transformed with respect to the phase
shift $\phi$ between the laser field components,
\begin{eqnarray} 
Y({\bf p},\phi) &=& Y_0({\bf p})+\vert Y_1({\bf p})
\vert e^{i\phi+i\Phi_1({\bf p})} \nonumber \\ & & \quad + \vert
Y_2({\bf p}) \vert e^{i2\phi+i\Phi_2({\bf p})} + \cdots + 
\mathrm{c.c.}\label{eq:series}
\end{eqnarray} 
The phases $\Phi_n ({\bf p})=
\mathrm{arg}\,Y_n({\bf p})$ describe
the phase lag of the change in the photoelectron yield as a function of the relative phase $\phi$, hence the name ``phase(s) of the
phase''. In the previous papers Ref.~\cite{Skruszewicz_PhysRevLett_2015,  Almajid_JPhysB_2017, Tulsky_PhysRevA_2018_POP}, only $\Phi_1({\bf p})$ was considered.  A representation of the series (\ref{eq:series}) is shown in Fig.~\ref{fig:fig1} (yields are normalized to a maximum value of 1 throughout the paper). The curves $Y_n({\bf p})=0$ in momentum space are very sensitive to the parameters of the system under consideration. Let us derive simple, analytical properties of $Y_n({\bf p})$ using the strong field approximation (SFA) \cite{Keldysh_1965, Faisal_1973, Reiss_1980}. Within the SFA, we can write the photoionization rate as
\begin{equation}\label{SFA_Yield_int} 
Y = \left\vert \int_{0}^{T}
P({\bf p},t) e^{{iS(t)}}dt \right\vert^2
\end{equation} 
where the prefactor $P({\bf p},t)$ (see the explicit form, e.g., in \cite{Popruzhenko_JPhysB_2014}) is slowly varying on the timescale of a laser period $T$, the fast oscillating function $S(t)$ in the exponent
\begin{equation}\label{Action} 
S(t)=\int_{0}^{t} \left[ \frac{({\bf
p}+{\bf A}(t'))^2}{2}+I_p\right]dt'
\end{equation}
is the classical action of the photoelectron with final
momentum ${\bf p}$, ${\bf A}(t)$ is the vector potential determining
the laser electric field ${\bf E}(t)=-\partial_t{\bf A}(t)$, and $I_p$
is the ionization potential.

We calculate the yield (\ref{SFA_Yield_int}) with exponential accuracy
using the saddle-point method (see, e.g. \cite{Milosevic_JModOpt_2006,
Popruzhenko_JPhysB_2014,Amini_RepProgP_2019} and references therein)
and obtain
\begin{equation}\label{SFA_Yield} Y({\bf p}, \phi) \sim \left\vert
\sum_s e^{iS(t_s)}\right\vert^2
\end{equation} where the complex times $t_s$ are solutions to the
saddle-point equation
\begin{equation}\label{SPeq} \frac{({\bf p}+{\bf A}(t_s))^2}{2}+I_p=0.
\end{equation} In the present paper, we aim at the case of two-color,
circularly polarized, counter-rotating fields described by a vector
potential of the form
\begin{equation}\label{A_t} \eqalign{ A_x(t)= A_0\left[\cos(a\omega
t)+\xi \cos(b\omega t+\phi)\right], \cr A_y(t)= A_0\left[\sin(a\omega
t)-\xi \sin(b\omega t+\phi)\right], }
\end{equation} and $A_z=0$, with the second component being weak
($\xi\ll 1$), and $a$, $b$ coprime, and $b>a$. For the sake of
simplicity, we assume an infinitely long pulse with a constant
envelope $A_0=E_0/a\omega$, where $E_0$ is the amplitude of the main
(i.e., strong) component of the laser field.

Introducing the Keldysh parameter $\gamma = \sqrt{2 I_p}/A_0$ and a
dimensionless momentum ${\bf q}={\bf p}/A_0$, one can rewrite
(\ref{SPeq}) as
\begin{eqnarray}\label{SPeq2} 0 &=& 1+\gamma^2+q^2+\xi^2
+2q\cos(a\omega t_s-\alpha) \\ && +2\xi[q\cos(b\omega
t_s+\alpha+\phi)+\cos((a+b)\omega t_s+\phi)]. \nonumber
\end{eqnarray} Here, $\alpha$ is the angle of photoelectron emission
in the $xy$ plane, i.e.,
\begin{equation} {\bf q} = q({\bf e}_x \cos\alpha + {\bf
e}_y\sin\alpha).
\end{equation} Most electrons will be emitted within the $xy$ plane as
long as non-dipole effects are negligible
\cite{Reiss_1980,PPT_1966}. Hence we do not consider final
photoelectron momenta out of the $xy$ plane in this work.

\begin{center}
	\begin{figure}
	\includegraphics[scale=0.088]{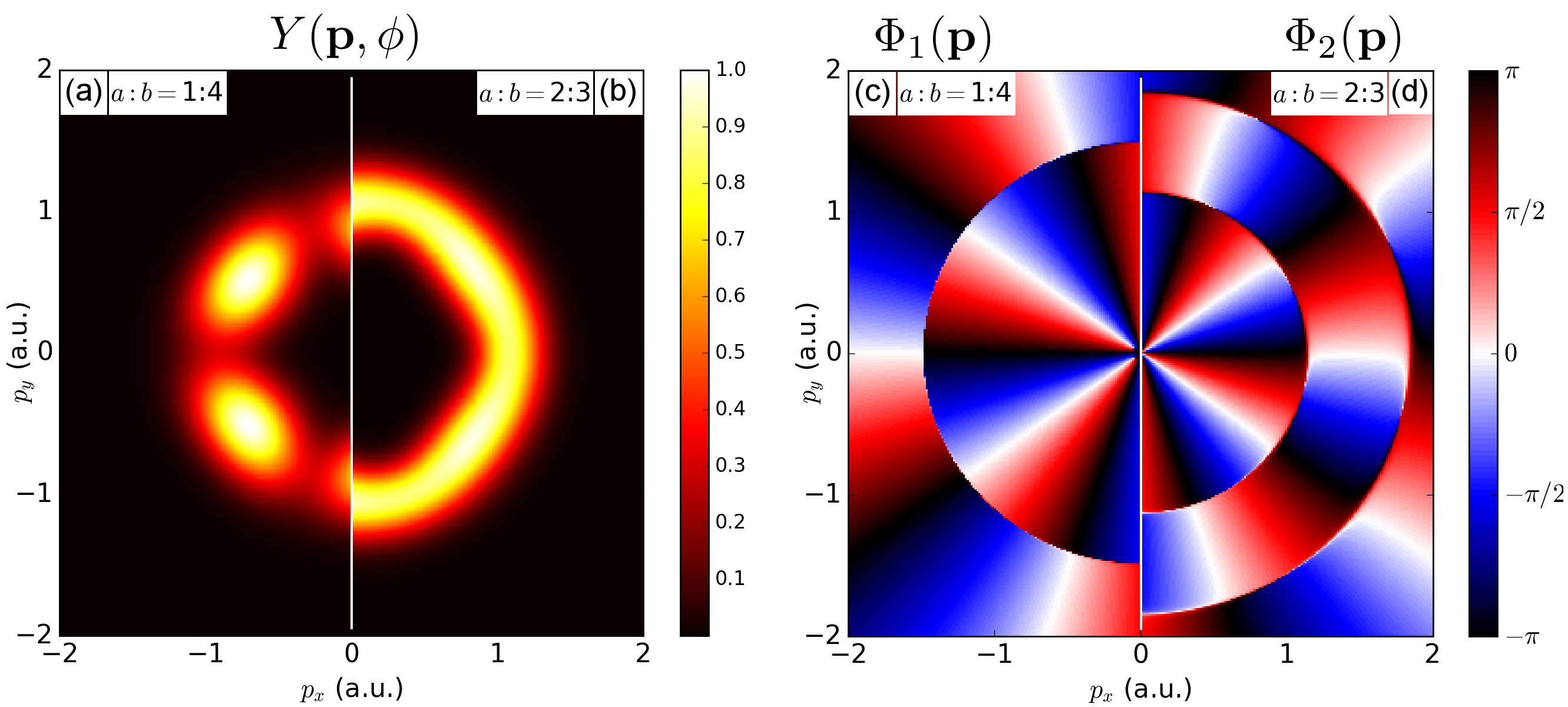}
	\caption{Left panel: PES obtained within the SFA for circularly polarized, counter-rotating (a) $\omega$-$4\omega$ and (b) $2\omega$-$3\omega$ fields. Intensity and wavelength of the main component of the laser $I = 2 \cdot10^{14}$ W/cm$^2$ and $\lambda_1 = 800$\,nm, $\xi=0.05$, $\phi=0$. Right panel: Corresponding PoP spectra: (c) $\Phi_1({\bf p})$ for $\omega$-$4\omega$  and (d) $\Phi_2({\bf p})$ for $2\omega$-$3\omega$.}
	\label{fig:fig2}
	\end{figure}
\end{center}

 In the monochromatic case ($\xi=0$), the emission is isotropic with
respect to the angle $\alpha$ \cite{PPT_1966} (see also
\cite{Smirnova_PhysRevA_2011}). Therefore, it is convenient to
introduce a new variable
\begin{equation} a\tau = a\omega t-\alpha.
\end{equation} In the two-color case, it is useful to introduce
additionally
\begin{equation}\label{theta} a\theta = (a+b)\alpha+a\phi.
\end{equation} One can then rewrite (\ref{SPeq2}) as
\begin{eqnarray}\label{SPeq3} 0 &=&
1+\gamma^2+q^2+\xi^2+2q\cos(a\tau_s) \\ && +2\xi
[q\cos(b\tau_s+\theta)+\cos((a+b)\tau_s+\theta)], \nonumber
\end{eqnarray} and the action (\ref{Action}) as
\begin{eqnarray} S(\tau_s) &=& \frac{A_0^2}{2a\omega} \bigg[
(1+\gamma^2+q^2+\xi^2) a \tau+2q\sin(a\tau) \nonumber \\ && +2\xi
\bigg( q\frac{a}{b}\sin(b\tau+\theta)\label{Action2} \\ && \qquad\quad
+\frac{a}{a+b}\sin((a+b)\tau+\theta) \bigg) \bigg]
\bigg\vert_{\tau=0}^{\tau=\tau_s}. \nonumber
\end{eqnarray} The yield only depends on $\alpha$ and $\phi$ in the
combination (\ref{theta}). The coefficient in front of $\alpha$ in
(\ref{theta}) is related to the $(a+b)$-fold symmetry of the electric
field and, hence, of the vector potential and the yield. As a
consequence, the yield has the property
\begin{equation}\label{properties} \eqalign{Y(q, \alpha, \phi) &=
Y\left(q, \alpha + \frac{a}{a+b}\phi, 0\right) \cr &= Y\left(q, 0,
\phi+\frac{a+b}{a}\alpha\right).}
\end{equation}

\subsection{General properties of the phase-of-the-phase} Now let us
pass over to the main quantity of interest, namely, the Fourier
components of the yield
\begin{equation}\label{Yn} Y_n(q, \alpha) = \frac{1}{2\pi} \int_0^{2\pi}Y(q, \alpha,
\phi)e^{-in\phi}d\phi.
\end{equation} Property (\ref{properties}) suggests to factor out the
$\alpha$-dependence of each $Y_n$,
\begin{equation} Y_n(q, \alpha) = Y_n(q, 0) e^{in \frac{a+b}{a}\alpha}.
\end{equation} 
This form allows to conclude that, due to the periodicity in $\alpha$, only harmonics which are multiples of $a$ are non-zero. In Fig.\ref{fig:fig2},  two 5-fold-symmetric configurations are shown, where in the $a:b=1:4$ case the first non-vanishing phase-dependent term is $Y_1({\bf p})$ while in the $a:b=2:3$ case it is $Y_2({\bf p})$. 
\begin{center}
	\begin{figure}
	\includegraphics[scale=0.085]{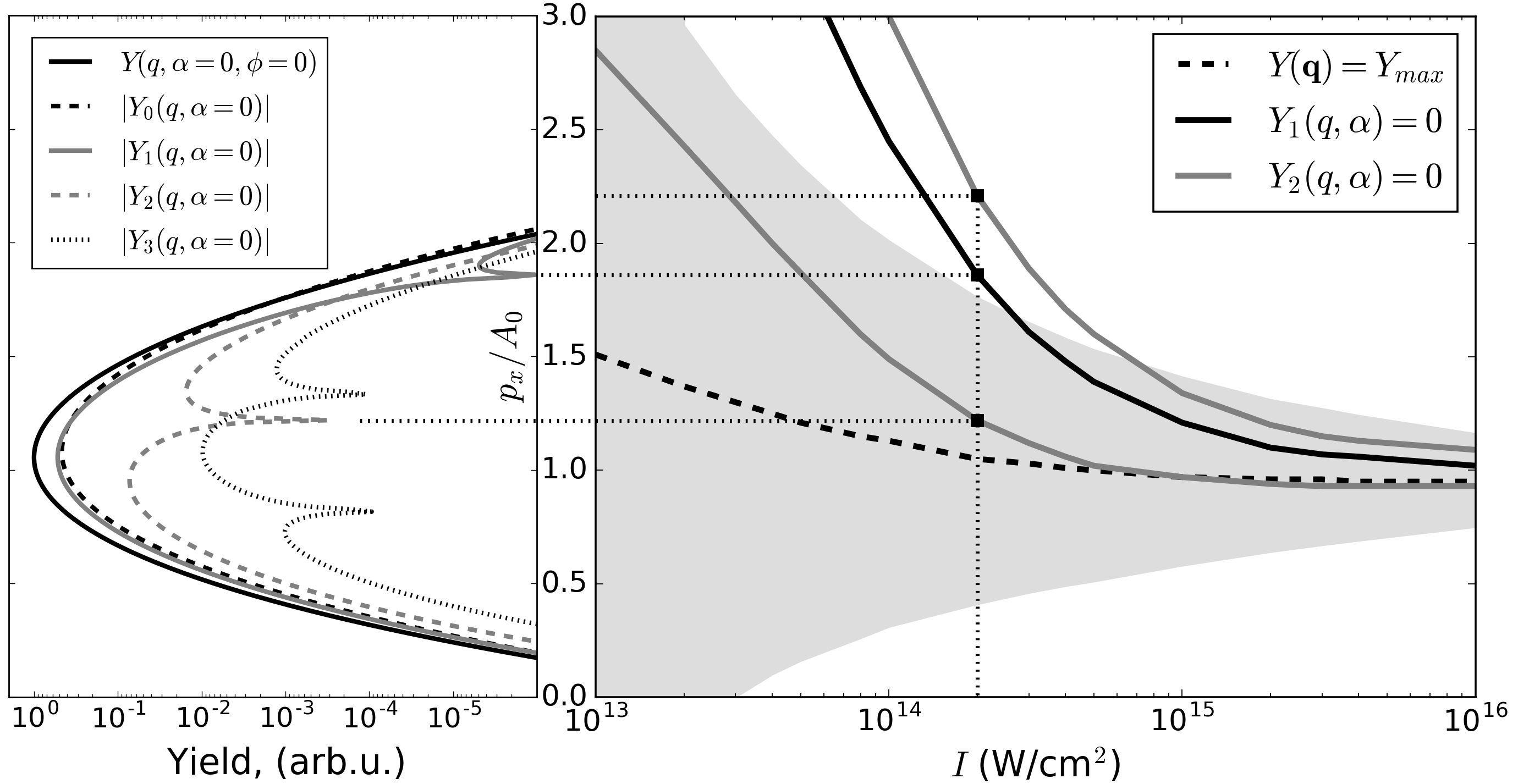}
	\caption{Left panel: Total yield $Y$ and absolute values of
several of the Fourier components $Y_n$ for $n=0$--$3$, taken along
the $q_x = p_x/A_0>0$ direction. All parameters are the same as in
Fig.\ref{fig:fig1}. Spiky minima correspond to momenta where the phase
flipping occurs and are indicated by squares on the right panel. Right
panel: Radial momenta of the phase flippings for the first (black
line) and the second (gray lines) harmonics of the Fourier expansion
for argon in a two-color $\omega$-$2\omega$ laser field with the laser main component's wavelength $\lambda_1 = 800$\,nm plotted versus the intensity of that laser field. The gray area represents a domain with a yield above $0.1\%$ of its maximum, and the dashed line indicates the position of this maximum.}
	\label{fig:fig3}
	\end{figure}
\end{center}
Another general property of the $Y_n$ is revealed if we consider the action and the saddle-point equation. One may easily
verify that if $t_s(p,\alpha=0,\phi)$ is a solution to (\ref{SPeq3})
for a given $\phi$, then $t_s(\alpha=0,-\phi) = 2\pi a -
t_s^*(p,\alpha=0,\phi)$ is a solution for $-\phi$. Next, inserting
these solutions into the action (\ref{Action2}) and evaluating the
imaginary part of it, one obtains the same value in both cases, i.e.,
$\Im S(\phi,t_s(\alpha=0,\phi)) = \Im
S(-\phi,t_s(\alpha=0,-\phi))$. Thus, the yield is symmetric in $\phi$
for $\alpha=0$. This allows to rewrite (\ref{Yn}) as
\begin{equation} 
Y_n(q, \alpha) = \frac{1}{\pi} \int_0^{\pi}Y(q, 0,\phi)\cos(n\phi)\, d\phi \, e^{in \frac{a+b}{a}\alpha}.
\end{equation} 
Finally, let us pass to the arguments $\Phi_n =
\mathrm{arg}\,Y_n$ of these Fourier components (i.e., to
phases-of-the-phase) that can now be written as
\begin{equation}\label{PoP} \Phi_n(q, \alpha) = n \frac{a+b}{a}\alpha
+ \pi \delta_n(q).
\end{equation}  Here, $\delta_n(q)$ is
related to the sign of the purely real $Y_n(q, 0)$ and can only have
values equal to 0 for $Y_n(q, 0)>0$ or 1 for $Y_n(q, 0)<0$ (without
loss of generality). Expression (\ref{PoP}) highlights an essential advantage
of the PoP technique related to circular, counter-rotating
fields: the only type of momentum-dependent features that appear in
the PoP spectra are circles centered around $q=0$ at which the phase
has a sharp flip of $\pi$. This makes the PoP flipping curves one
dimension simpler compared to the case of linear+linear two-color
fields and allows to represent their behavior for a broad variety of
parameters within a single plot (see the example shown in
Fig.\ref{fig:fig3}), thus making PoP a possibly handy
laser-calibration technique.

\subsection{PoP flipping of the higher Fourier components} 
Electron emission in circularly polarized fields at intensities above
$10^{14}$ W/cm$^2$ and $I_p/\omega\gg 1$ has the highest probability
for radial momentum $q = 1$, i.e., $p = A_0 = E_0/a\omega$
\cite{PPT_1966}. In Ref.~\cite{Tulsky_PhysRevA_2018_POP}, the position of
the flipping of the PoP of the first Fourier component $\Phi_1$ for
$\omega$-$2\omega$ laser fields (i.e., $a=1,b=2$ in the current
notation) has been analytically described. It appeared that the circle
of the $\Phi_1$ flipping is in general located in the momentum space
region far beyond the peak of the ionization probability where the
yield might be too low in actual measurements. Here we show that in
such cases the PoP technique using higher harmonics of the Fourier
expansion should be more practicable from the experimental point of
view. In particular, for the laser parameters considered above,
$\Phi_2$ flips close to the maximum yield (see example in
Fig.\ref{fig:fig3}) and, thus, should be easier to observe than the
flipping of $\Phi_1$.

\section{PoP-based exclusion of incoherent scattering}
\begin{center}
	\begin{figure}
	\includegraphics[scale=0.042 ]{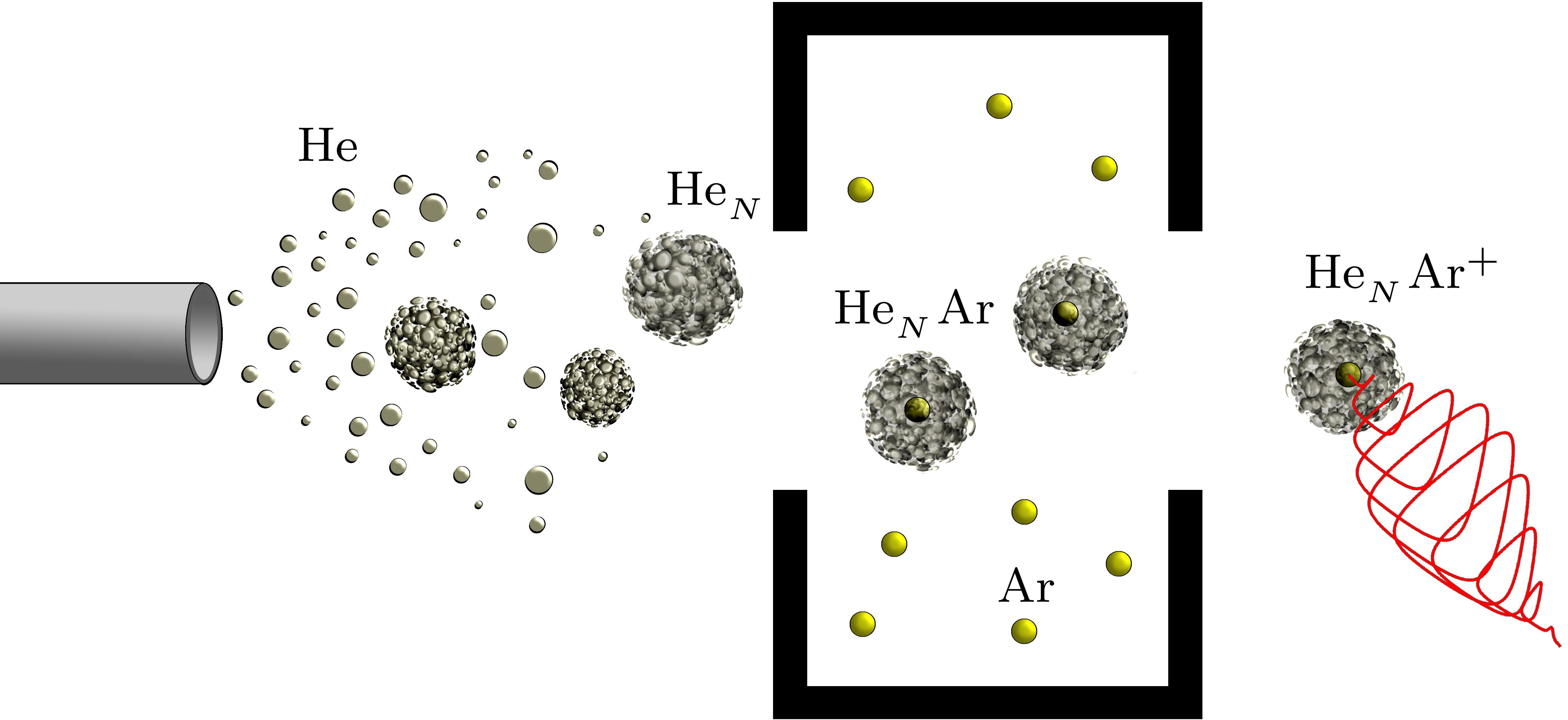}
	\caption{Sketch of the setup used in our calculations. Each
helium droplet of size $N$ captures a single argon atom, which is then
ionized by a $\omega$-$2\omega$ laser field.}
	\label{fig:fig4}
	\end{figure}
\end{center} 
\begin{center}
	\begin{figure}
	\includegraphics[scale=0.098]{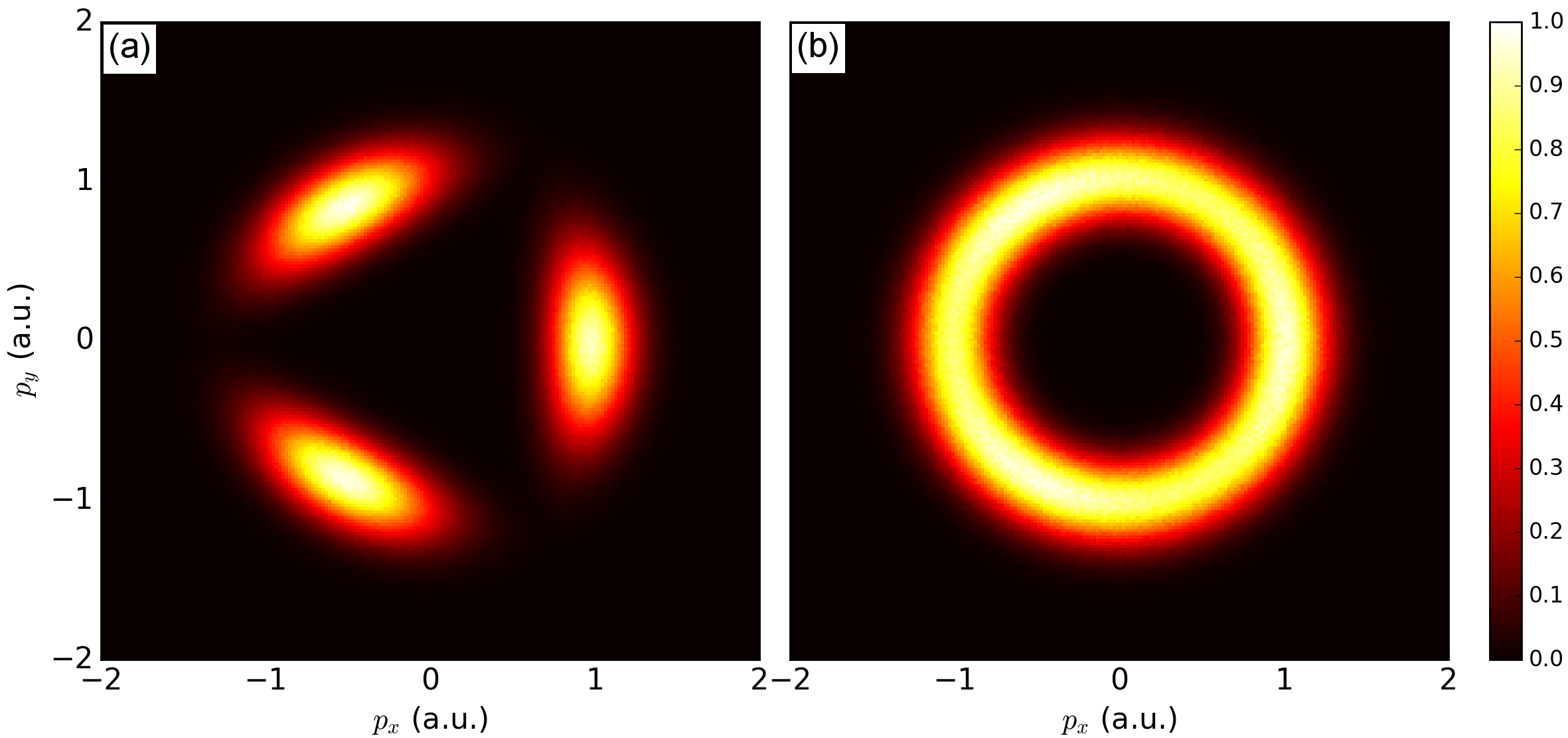}
	\caption{Comparison of PES obtained within the SFA for bare (a) and embedded (b) argon in the circularly polarized, counter-rotating
$\omega$-$2\omega$ field. In (b), helium droplets  of size $N=3
\cdot 10^3$ are considered, which leads to a blurring of the PES due to elastic e-He scattering. Intensity and wavelength of the main component of the
laser $I = 2 \cdot10^{14}$ W/cm$^2$ and $\lambda_1 = 800$\,nm,
$\xi=0.05$, $\phi=0$. $N_e =2 \cdot 10^7$ photoelectrons are simulated.}
	\label{fig:fig5}
	\end{figure}
\end{center}

In order to demonstrate another practical application of the PoP technique, we introduce a model describing  the PES generated by ionization of atoms in a complex environment. In particular, we consider helium nanodroplets \cite{Barranco_JLowTemp_2006}. They are widely used as finite spectroscopic matrices for embedded particles, called dopants \cite{Toennies_2004, Stienkemeier_JPhysB_2006, TigPCCP07, Yang_ChemSocRev_2013, Mudrich_IntRevPhysChem_2014}. 
The ultralow temperature of the droplets, the weak electronic interaction between dopant and helium, and the optical transparency up to about 20\,eV  allows for almost undisturbed high resolution spectroscopy in the microwave, IR and UV spectral range of  targets varying in complexity
from simple atoms \cite{Wang_JPhysChemA_2008} to large
biomolecules \cite{Wang_JPhysChemA_2008,Dong_Science_2002,Lehmann2065}. In contrast, photoelectrons emitted by the dopant might scatter on the surrounding helium atoms before leaving the droplet. In that way, an incoherent electron yield is added to the PES since the phase information is almost lost in the scattering. This incoherent contribution might mask the interesting, laser-coherent features in the ordinary PES while PoP spectra still show them, as will be demonstrated in the following.

In our computer experiment, we use argon as the dopant. The laser parameters considered are as follows: two counter-rotating, circularly polarized components with wavelengths $800$ nm and $400$ nm ($a=1,b=2$), intensities are $I_1 = 2 \cdot 10^{14}$ W/cm$^2$ and $I_2 = 2 \cdot 10^{12}$ W/cm$^2$ ($\xi=0.05$). A sketch of the experimental setup to produce single-atom-doped helium droplets is shown in Fig.\ref{fig:fig4}. 
We consider droplets of size $D \approx 60 $\,\AA,  that is $N \approx 3 \cdot 10^3$ atoms. At 20\,bar and for a 5$\mu$m diameter gas exit, nozzle temperatures of about $16$\,K have to be established~\cite{KelJCP19}. The partial pressure of argon gas in the pick-up region is adjusted to conditions such that only a single Ar atom is present in each droplet on average. DFT calculations show that rare gas impurities reside near the center of the droplet \cite{Coppens2017a,Coppens2017b}.

The photoelectrons have a significant probability of multiple scattering on the enclosing neutral helium atoms before leaving the droplet and flying towards the detector. As the mean distance between the He is larger than their size, we treat each scattering within a single-atom approach. Due to the significant difference between the ionization potentials of argon ($15.8$\,eV) and helium ($24.6$\,eV), we neglect the ionization probability of the latter and consider photoelectrons produced from argon only.  For the sake of simplicity, we reduce our problem to the two-dimensional case, taking into account the polarization plane only, since most of the photoelectrons produced from an atom in a circularly polarized laser field have their momenta within the polarization plane. As a result, most of the coherent signal is in the momentum distribution in this plane. As typical kinetic energies of photoelectrons produced from argon with the chosen laser intensity are too low to ionize or excite helium, we only take elastic scattering into account. Moreover, most electrons have an energy close to the average $\langle p^2\rangle /2=15.4$\,eV so that we neglect the dependence of the cross section on the energy of the scattered electrons. 
The total cross section $\sigma$ and the angular distribution $d\sigma / d\Omega$ at $p^2/2=15$\,eV are taken from~\cite{ioffe_elastic}.
  For $N_{\phi} = 20$ phases $\phi \in [0,2\pi[$ (i.e., 10 sampling  points per period of $Y_2$) we simulate a finite number $N_e$ of ionization events with the initial momentum distribution taken from the SFA. Then we simulate their scattering with the probability of traveling a distance $s$ before a new scattering event as
\begin{equation}\label{w_s} 
w(s)= 1-e^{ -n \sigma s} 
\end{equation} 
where the concentration of helium atoms $n$ is assumed to be constant all over the droplet. Thus, $s$ is chosen randomly as $s=-(n \sigma)^{-1} \log R $ where $R$ is a random number between 0 to 1. If $s$ is smaller than the distance to the boundary of the droplet, the scattering occurs. The angle of scattering $\vartheta$ is then chosen randomly in a range from $-\pi$ to $\pi$ according to the angular distribution $d\sigma / d\Omega $. The total number of scattering events for each initial electron is not restricted.
\begin{center}
	\begin{figure}
	\includegraphics[scale=0.158]{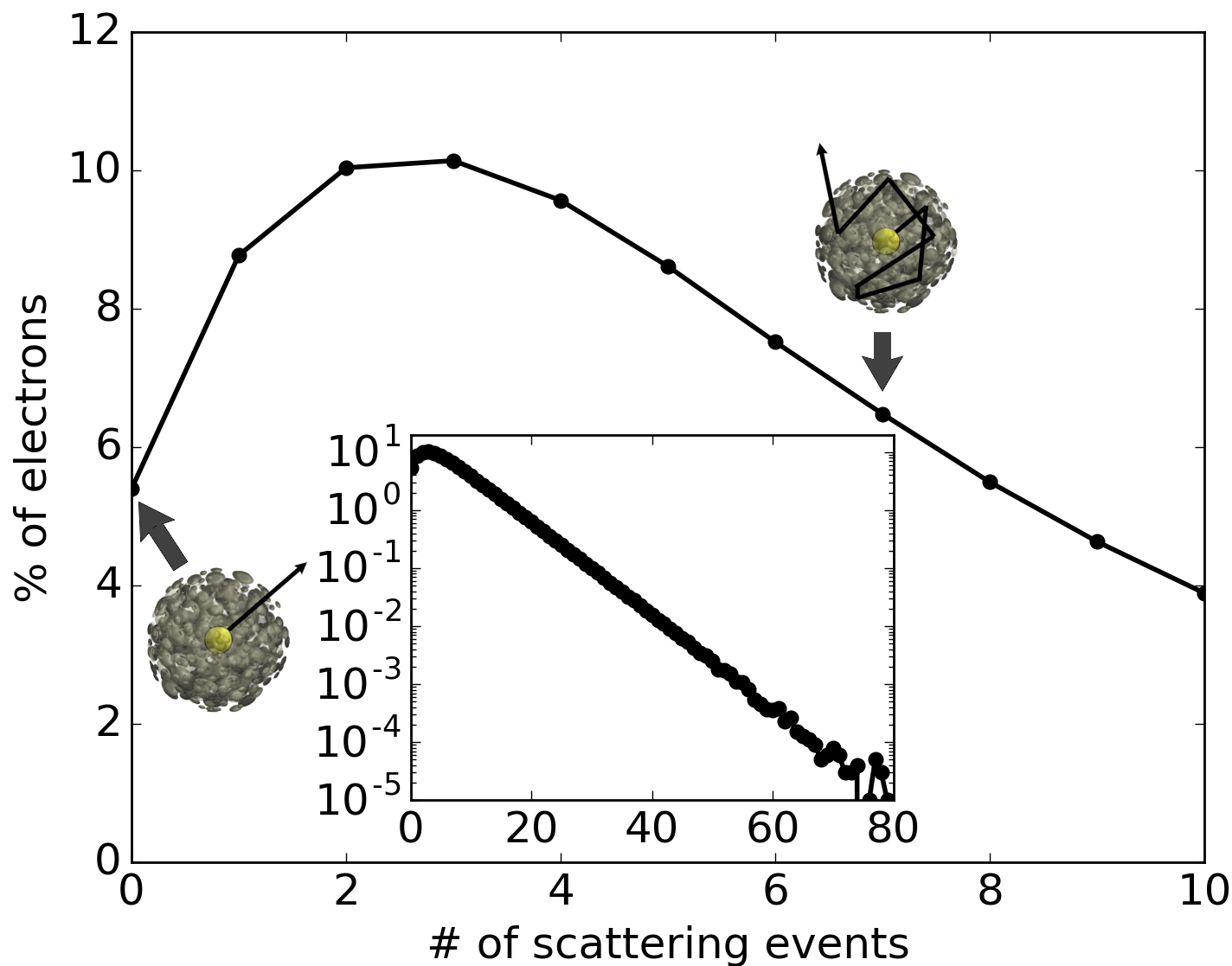}
	\caption{Percentage of the total number of photoelectrons
          $N_e=2 \cdot 10^7$ that experienced a certain number of
          elastic scattering events before leaving the helium
          droplet. The inset shows the distribution logarithmically
          and over a wider range of scattering events.}
	\label{fig:fig6}
	\end{figure}
\end{center}

\begin{center}
	\begin{figure}
	\includegraphics[scale=0.098]{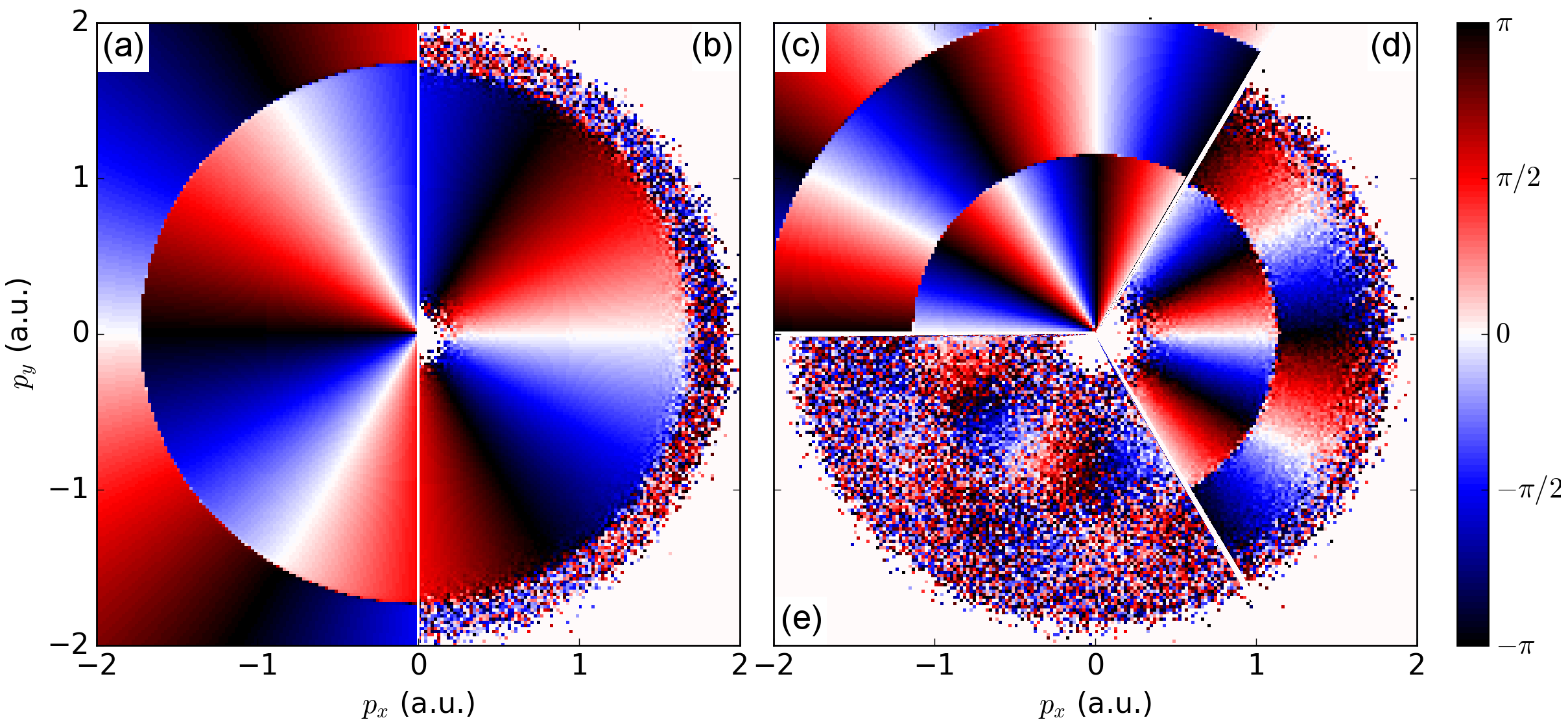}
	\caption{Phase-of-the-phase (a) $\Phi_1$ predicted by the
SFA. (b) $\Phi_1$ obtained with a finite number of $N_e=2 \cdot 10^8$
photoelectrons for each phase $\phi$. (c) $\Phi_2$ predicted by the
SFA. (d) $\Phi_2$ obtained with a finite number of $N_e=2 \cdot 10^7$
photoelectrons for each phase $\phi$. (e) Same as (d) but with
elastic scattering on neutral helium atoms in the droplet taken into
account. In each case $N_{\phi} = 20$ phases $\phi \in [0,2\pi[$ have
been simulated.}
	\label{fig:fig7}
	\end{figure}
\end{center}

An example of a PES obtained with the SFA is shown in
Fig.\ref{fig:fig5}(a). If an atom is trapped inside a helium droplet
the PES is significantly distorted because of electron-He scattering
(see Fig.\ref{fig:fig5}(b)). The 3-fold symmetric laser-coherent part
of the total signal is now deeply suppressed by an almost circular PES
of multiply scattered electrons. The percentage of electrons as a
function of the number of scattering events in Fig.\ref{fig:fig6}
shows that only about $6\%$ of the laser-coherent signal survives. After
having simulated a set of PES for different relative phases $\phi$, we
apply the PoP technique and obtain the phase flipping curves (see
Fig.\ref{fig:fig7}). Due to the low ionization probability of
photoelectrons with high momenta, the flipping curve for $\Phi_1$
(which is at $p=1.76$\,a.u.) is only visible if a huge number of
ionization events is considered ($N_e \geq 2\cdot 10^8$ for a grid
with $dp=0.02$\,a.u.) even without scattering taken into
account. However, as previously noted, instead of increasing the
number of particles (i.e., the measurement time in the experiment) one
may simply consider $\Phi_2$, the phase of the second harmonic in the
Fourier expansion. As was predicted in Fig.\ref{fig:fig3}, it
requires a significantly smaller $N_e$ for the phase-flipping to
remain visible, even when the spectra are distorted by scattering, as
shown in Fig.\ref{fig:fig7}(e).

\section{Conclusions}

To wind up, we showed that a slight extension of the phase-of-the-phase
technique applied to a two-color, circularly polarized and
counter-rotating intense laser field can significantly improve its
practicability: considering higher terms in the Fourier series of the
photoelectron yield, phase-of-the-phase spectra with flipping curves
that are located in regions of high photoelectron yield can be
obtained.  Being based on the observation of sharp and
intensity-sensitive features, phase-of-the-phase spectroscopy could be
an instrument for accurately tuning and benchmarking the laser
intensity.

Secondly, we performed a simulation of photoelectrons originating from
the ionization of a single argon atom inside a helium droplet by a
$\omega$-$2\omega$ laser field. The electrons may scatter on their way
to the detector so that laser-coherent features in the photoelectron
spectra are obscured. Simulating the spectra with a finite number of
electrons mimics the effect of a finite measurement time and allowed
to treat the effect of (multiple) scattering events on the electrons'
pathways to the detector classically and statistically. Finally, we
showed that laser-coherent features survive in the second-harmonic
phase-of-the-phase spectrum calculated in such a manner.

\section*{Acknowledgments} This work was supported by the projects BA~2190/10, TI~210/7, and TI~210/8 of the German Science Foundation (DFG).

\section*{References} \bibliography{biblio}

\end{document}